\documentstyle[aps]{revtex}

\def\ba{\begin{eqnarray}}
\def\ea{\end{eqnarray}}

\begin{document}

\title{Thermodynamic constraint on the  primordial black hole
formation in the radiation dominated epoch}

\author{Hyun Kyu Lee\footnote{e-mail: hklee@hepth.hanyang.ac.kr}}

\address{Department of Physics, Hanyang University, Seoul 133-791, Korea \\
and \\ Asia Pacific Center for Theoretical Physics, Pohang
790-784, Korea}

\date{\today}
\maketitle

\begin{abstract}
It has been suggested that the overdense region as a result of
inhomogeneities in the early Universe  would have undergone a
collapse into the primordial black holes(PBH).  In this work, we
discuss  a possible constraint on the PBH formation in the
radiation dominated epoch  by imposing   the generalized second
law of thermodynamics in the context of spherically collapsing
scenario. It is found that both the critical temperature $T_c$
over which the formation of PBH is not possible and  the lower
bound on the mass of PBH  depend on the number of degrees of
freedom at the time of PBH formation. In the standard model,  one
can show that the lower bound on the mass of PBH  known in the
literature, of order Planck mass, is consistent with the
thermodynamic constraint constructed in this work. We also pointed
out the possibility that the critical temperature(lower bound on
PBH mass) can be lowered(increased) provided  the number of
relativistic degrees of freedom of the Universe is increasing
substantially beyond the standard model. 
\end{abstract}

There are many observational indications , for example, the
existence of the galaxies and the fluctuations observed in CMBR,
which imply   that the early Universe must have been
inhomogeneous. It has been suggested that the overdense region as
a result of inhomogeneities in the early Universe  underwent into
the primordial black holes(PBH) \cite{zeldovich}\cite{hawking}.
The existence of PBH has been considered to have interesting
cosmological consequences, for example, on the cosmic microwave
background radiation, primordial nucleosynthesis and   dark matter
candidate. The mass distribution of PBH  can  provide a useful
information on the spectrum of density fluctuations in the early
Universe. Hence the formation  as well as the evolution of PBH
have been interesting subjects of investigations
\cite{carr}\cite{jedamzik}\cite{horvath}. Since PBH is an issue in
the early Universe, the discussion  depends on the theory of
gravity, which might be different from that of Einstein. For
example, the formation and evolution of PBH has been discussed
also in scalar-tensor theories of gravity \cite{barrowcarr}.

One of the remarkable developments in black hole physics is the
thermodynamic understanding of a black
hole\cite{bekenstein}\cite{bch}. The entropy of a black hole is
proportional to the surface area of horizon and the temperature to
the surface gravity.  Then the second law of thermodynamics can be
stated in a generalized way  that the total entropy of black hole
plus external Universe never decreases.   Hence it is interesting
to see whether the generalized second law of thermodynamics can
impose  any constraint on the physical processes in which black
holes are involved. The purpose of this work is to discuss how the
formation of PBH can be constrained by imposing the generalized
second law of thermodynamics, which is believed to be valid during
the spherically collapsing process.

The evolution of  the overdense gravity-dominated-regions is
supposed to be described by the Friedman-Robertson-Walker(FRW)
metric. Since overdense region is characterized with the spatial
curvature,  $k
> 0 $,  in the background of expanding Universe, the scale factor $R$ is
decreasing  to form a black hole at a later stage.  In this
spherically-collapsing scenario the entropy in a comoving volume
is conserved  without any loss or any net creation up to a black
hole threshold . It is normally assumed  that PBH mass is of order
horizon mass in the early Universe\cite{carr}, which is supposed
to be radiation dominated with temperature $T$. Then the PBH mass
has temperature dependence as $M_{pbh} \sim T^{-2}$ and   the
entropy of PBH according to the black hole thermodynamics behaves
as $ S_{pbh} \sim T^{-4}$.  Since the matter in the particle
horizon is a kind of progenitor of PBH, it is interesting to
compare $S_{pbh}$ with the entropy enclosed in a particle horizon,
$ S_{hor} \sim T^{-3}$. One can see that there must be a
cross-over temperature $T_c $ above which PBH entropy is smaller
than that of its progenitor.   It implies that the the formation
of PBH at $T>T_c$ may not be possible without violating the second
law of thermodynamics. In this  work, we discuss the possible mass
limit of PBH using an adiabatic and spherical collapsing  model of
overdense region into a black hole. The FRW metric is used both
for collapsing region and expanding background Universe, which are
assumed to be  radiation dominated.

It is well known that a spherical collapse of matter results in a
black hole if the repulsion due to pressure  is not sufficient
against gravitational collapse\cite{os}.  According to Birkhoff
theorem, during the spherical collapse there is no observable
change outside the collapsing matter. For example, there is no
gravitational nor electromagnetic  waves which might otherwise
carry out   the energy away  and create the entropy. The evolution
in FRW metric is characterized by the scale factor, $R$, defined
by
\ba
ds^2 = -dt^2 + R^2(\frac{dr^2}{1-\kappa r^2} + r^2
d\Omega^2)\label{frw}, \ea where $\kappa$ is positive for a
collapsing matter. The evolution of overdense region in the early
Universe is supposed be described by eq.(\ref{frw}) with $\kappa$
identified as the perturbed total energy per unit mass\footnote{
In the work of Oppenheimer and Snyder\cite{os}, $\kappa $ is given
by $ \kappa = \frac{8\pi}{3} G R(0)^2\rho(0) ~$, where $R(0)$ and
$\rho(0)$ are the scale factor  and the mass density of
pressure-less dust when collapse starts.}.

The early Universe is  assumed to be  in  radiation dominated era
in  thermal equilibrium (with temperature $T$), where the energy
$U$ can be expressed   in terms of energy density $\rho$ and
volume : $ U = \rho(T) V$. Using the thermodynamic identity
\ba
T dS = dU + p dV \ea one can get
\ba
dp &=& \frac{(\rho +p)}{T} dT, \label{pt} \ea and
\ba
 dS &=& d \frac{(\rho +p)V}{T}. \label{st} \ea  Using   eq.(\ref{pt})
 and the energy
conservation for a comoving volume,
\ba
d ((\rho + p)R^3) = R^3 dp \label{pr},  \ea   we get
\ba
d[R^3(\rho+p)/T] =0. \ea  Compared with eq.(\ref{st}), it  shows
that the entropy in a comoving volume is conserved\cite{weinberg}.
It implies that the entropy of collapsing matter which are fixed
in a comoving frame is conserved during the collapsing process up
to black hole threshold.

When it becomes  black hole then the entropy is determined by the
completely  different reasoning regardless of whatever the amount
of entropy has been carried in.  The entropy of black hole is
determined  by the surface area of the horizon and for the
Schwarzschild black hole it is given by
\ba
S_{bh} =  \frac{4\pi G}{\hbar c} M_{bh}^2 \ea in the unit of
Boltzmann constant $k_B$. There have been  many studies  in
understanding the microscopic nature of black hole entropy e.g.,
using membrane paradigm\cite{tpm} and recently in string
theory\cite{bhs}. Hence the discontinuity in the entropy between
the advected one and the black hole entropy is naturally expected
during collapsing procedure.   As an example, if one compare the
entropy of the Sun and the entropy of black hole of the same mass,
the difference of entropy is quite large:
\ba
S_{\odot} \sim 10^{54} \ll S_{bh} \sim 10^{77}
\label{bhentropy}\ea
 We consider the
issue how this discontinuity can be understood microscopically
near at the black hole threshold is a separated problem beyond of
this work.

We assume that the second law of thermodynamics should be obeyed
during the collapsing procedure whatever  the microscopic
understanding of the black hole entropy is turned out to be. This
gives us a condition that the black hole entropy should not be
smaller than the advected entropy,
\ba
S_{bh} \geq S_{adv} , \ea since there is no out-going entropy in
this spherically collapsing scenario. This gives  possible  limits
on the formation temperature and  the mass of PBH which is being
described below. It is interesting to note that  the cosmological
holographic constraints on the entropy\cite{kaloper}  have been
discussed  in the similar context: the  entropy of matter inside a
black hole should be smaller than the black hole entropy. In
ref.(\cite{kaloper}) the discussion is given using order of
magnitude estimations. In this work, however, a detailed
calculation will be given by taking account of the matter content
of a underlying theory and also the dynamics of collapsing
procedure.

The mass of PBH is estimated to be of order $M_{hor}$ in the early
Universe with equation of states $p = \gamma \rho $.  It is
because PBH mass must be bigger than the Jeans mass $ \sim
\gamma^2 M_{hor}$ but smaller than the horizon mass
itself\cite{carr}. It is fair to assume that the entropy being
advected along the collapsing matter is of order $S_{hor}$. As an
approximation for order-of-magnitude estimations we can simply put
\ba
M_{pbh} = M_{hor}, ~~~ S_{adv} = S_{hor}. \ea  Since there is no
change of entropy in background expanding Universe,  the condition
from the second law of thermodynamics that the entropy of final
black hole should not be smaller than the advected entropy can be
written as
\ba
S_{hor} \leq  4\pi (\frac{M_{hor}}{m_{pl}})^2, \ea where  $m_{pl}
= \sqrt{\hbar c/ G}(= 2.18 \times 10^{-5}$g) is the Planck mass.

The energy density and entropy density of the radiation dominated
Universe are given by
\ba
\rho = \frac{\pi^2}{30} g_* T^4, ~~ s = \frac{2 \pi^2}{45} g_*
T^3, \ea where $g_*$ is determined by the relativistic degrees of
freedom at the temperature $T$ of the  Universe\cite{weinberg}.
 Then mass
and entropy enclosed in the horizon  can be obtained as
\ba
S_{hor} = \frac{0.050}{\sqrt{g_*}}(\frac{m_{pl}}{ T})^3,
\label{shor}\\ M_{hor}=\frac{0.038}{\sqrt{g_*}}
(\frac{m_{pl}}{T})^2 m_{pl}.
 \ea
The horizon radius\footnote{The horizon radius in this work is
defined by a half of causal distance, $d_H$, since the matters in
a distance of causal distance only are relevant in this
discussion\cite{weinberg}.} is given by
\ba
r_{hor} = \frac{0.30}{\sqrt{g_*}}(\frac{m_{pl}c^2}{k_B T})^2
l_{pl}
,\label{rhor}\ea where the Planck length is defined by
$l_{pl}= \hbar / m_{pl} ~ c = \sqrt{\frac{G \hbar}{c^3}}$.

Since $M_{pbh}$ is identified with $M_{hor}$ in this work, the
radius of PBH horizon, $r_{pbh}$, is given by
\ba
r_{pbh} &=& 2 M_{hor} G/ c^2 = 2(\frac{ M_{hor}}{m_{pl}}) l_{pl}
 .\ea
Compared with the horizon radius,  one can see that the event
horizon of PBH, $r_{pbh}$,   is smaller than horizon radius,
$r_{hor}$,
\ba
r_{pbh} = .25 ~ r_{hor}, \ea which is consistent with the
collapsing scenario into PBH.

Now  the entropy of PBH is given by
\ba
 S_{pbh}
 = \frac{0.018}{g_*}
 (\frac{m_{pl}c^2}{k_BT})^4.
\ea We can see
 that  PBH entropy is decreasing faster with temperature
than the advected entropy, eq.(\ref{shor}), and there is a
cross-over temperature\cite{nsl}, $T_c$,  beyond which the entropy
of PBH is smaller than that of advected matter.  Hence the
critical temperature is given by
\ba
k_B T_c = \frac{0.36}{\sqrt{g_*}} m_{pl} c^2 .
\ea
 PBH may form
only when the temperature of the Universe drops below $T_c$. And
the critical mass $m_{pbh} \geq M^c_{hor}$ is given by
\ba
M^c_{hor}  = 0.29 \sqrt{g_*} m_{pl}. \label{mc}\ea

It is interesting to note that  within a factor of order unity it
is consistent with condition that the Compton wavelength of PBH, $
\lambda_{pbh} = \frac{\hbar}{M_{hor}c}$, is within the particle
horizon:
\ba
 \lambda_{pbh} ~ <
  ~ r_{hor}= 7.92 ~ \frac{M_{hor}}{m_{pl}} \lambda_{pbh}
 \ea

One can see that $T_c$ depends on  the  number of degrees of
freedom:  when we  take larger value of $g_*$,  we get    lower
critical temperature and the lower bound on PBH mass becomes
higher. With $g_* = 106.75$ in the standard model and using the
presently observed gravitational constant $G_0=6.67 \times
10^{-11} m^3 kg^{-1}s^{-2}$, we get
\ba
k_B T_c = 3.5 \times 10^{-2} m_{pl}c^2. \ea and $M^c_{hor}$ in
eq.(\ref{mc}) is given by
\ba
M^c_{hor} = 3 ~ m_{pl}, \ea which is similar to the lower bound
conventionally quoted in the literature.
 This result shows that  the lower
bound on PBH mass discussed in the other context using order of
magnitude estimations is in fact consistent with the
thermodynamics of black hole.

To complete the numerical estimations, the critical horizon mass,
eq.(\ref{mc}) is compared to the horizon mass at Planck time, $
t_{pl} \equiv l_{pl}/c( = \frac{\hbar}{m_{pl}c^2})$, given by
\ba
M_{hor}(t_{pl}) =  0.069 m_{pl}. \ea   One can see that it is
bigger by two-orders of magnitude than the horizon mass at Planck
time. Similar estimation shows that  the density at the critical
temperature is much smaller than the density at Planck
temperature:
\ba
\rho(T_c) = \frac{0.18}{g_*} \rho(T_{pl}), \ea where the
temperature at Planck time is given by $T_{pl} = 0.55 ~ m_{pl}/
g*^{1/4}$ and the energy density at Planck time($t_{pl}$) by
$\rho(T_{pl}) = \frac{3}{32\pi} \frac{m_{pl}}{l_{pl}^3}$ in the
standard model.

 It should be noted
that the parameters, $g_*$ and $G$, in determining the critical
temperature can have different values from the standard model
depending on the new physics at such high temperature stage of the
Universe. It is possible that up to this temperature there might
be more degree of freedom than the standard model with new
physics, e.g., super-symmetry. Then  the critical temperature
might be lowered. On the other hand the possibility of varying
gravitational constant at the early stage of Universe arises
naturally in various scalar tensor theories of gravity. At the
earlier stage of Universe a strong time variation is possible
according to the generalized scalar tensor theories. However the
strong variation is only possible for the vacuum dominated era. In
the radiation dominated era which follows vacuum dominated era, no
change of gravitational constant is expected. Therefore for PBH's
formed in the radiation dominated era, the change the critical
temperature due to the variation of the gravitational constant
from the present value of $G_0$ is not expected to be substantial
except the case in which the overdense region carries
``gravitational memory"\cite{barrowcarr}. It is because although
the change of gravitational constant is possible in the matter
dominated era following the radiation dominated era it is
restricted by the nucleosyntheis analysis.  Up to back in
nucleosynthesis era, only a small variation of the gravitational
constant  from the presently observed value of $G_0$ is found to
be consistent with the observed primordial abundance\cite{nucl}.

In summary it is  demonstrated that there is a possible constraint
on the PBH formation imposed by the second law of thermodynamics,
if it is formed by  spherical collapse of overdense region during
radiation-dominated era\footnote{It should be noted that PBH can
be produced in the vacuum dominated era, which is proposed for the
earlier phases of the Universe preceding the radiation dominated
era. Then the constraint discussed above cannot be strictly
applied to the black holes formed in this epoch, since no entropy
is associated with a vacuum energy such that black hole entropy
always wins.}
 in the background of expanding Universe.  It is assumed that PBHs
 can form with a horizon mass and the entropy of comoving volume
 in the collapsing matter and expanding background Universe
 is conserved up to black hole threshold. In the radiation
 dominated Universe, the event horizon of PBH is always smaller
 than the particle horizon.

 By  comparing  the entropy of the black
hole at the later stage  with that of advected entropy of
collapsing matter, one can find a constraint that there is a
critical temperature $T_c$ above which no PBH can be formed
without violating the second law of thermodynamics.
 The critical temperature appears to depend on the physics in the
 earlier epoch   than GUT scale: the number of degrees of freedom
 at the time of PBH formation.
 In the standard model, we can demonstrate explicitly
that the thermodynamical lower bound of PBH mass as well as the
critical temperature is of order Planck scale. This result shows
that  the lower bound on PBH mass discussed in the other context
using order of magnitude estimations is in fact consistent with
the thermodynamics of black hole. However beyond the standard
model there is a possibility that the critical temperature can be
lowered if the number of relativistic degrees of freedom of the
Universe  is increasing substantially.

It is  also found  that the entropy condition on PBH formation  is
consistent with the condition that the Compton wavelength should
be within the particle horizon for the semiclassical treatment of
PBH in this work. Recently there is an interesting work on the
entropy of a black hole surrounded by a thermal
atmosphere\cite{page}, in which the black hole entropy is shown to
be   less than the classical one adopted in this work. Assuming
that it can be applied  also for the collapsing procedure, one can
see that  the effect of the background,  the expanding Universe in
thermal equilibrium, does not relax the constraint on PBH
formation but makes  the constraint more stringent by lowering the
critical temperature.

If the critical temperature becomes substantially lower than
Planck temperature such that failed-collapse into PBH might be
abundant, there is a question on what would be the later stage of
collapse if PBH cannot be formed. There should be a physical
process in which a substantial part of energy is transported
outward with entropy creation, before the last envelope of
collapsing matter reaches `would-be event horizon'. It may be in
the form of radiation. Although it depends also on the presence of
`would-be event horizon', the characteristics may not be the same
as Hawking radiation, which is believed to be valid only for PBH
formed at $T < T_c$. Therefore  one can suspect a radiation from a
collapsing object at $T > T_c$  with different characteristics
from Hawking radiation. Hence if it is possible to implement these
effects of $T_c$  into the observational data and/or in evolution
models of early Universe, it might be useful in  accessing   new
physics around/above GUT scale, e.g., supersymmetry and
scalar-tensor theory of gravity.

\acknowledgments
 The author would like to thank Hongsu Kim and
Sang-Pyo Kim for elaborating discussions.  The author also thanks
anonymous referee for valuable comments in clarifying several
points in this work.
 This work was supported by grant No. 1999-2-11200-003-5 from the
Korea Science \& Engineering Foundation.


\begin{thebibliography}{}
\bibitem{nsl} Similar analysis is possible also for a spherical collapse of
neutron-star-like object.  It is supposed to be a degenerated
neutron matter with temperature much smaller than the Fermi
temperature, $T \ll T_F$.  The entropy is then proportional to the
number of neutrons $ S_{NS} \sim  N $, up to a factor which
depending on  $\frac{T}{T_F}$ which is assumed not to change
substantially. Then the entropy of neutron star matter can be
approximated as $ S_{NS} \sim M_{NS}$.  The entropy of
string-matter is also known to be approximately proportional to
the energy\cite{bhs}. As in PBH formation, when it collapses into
a black  hole, the entropy of black hole  is given by $ S_{BH} =
\frac{4\pi}{\hbar} M_{NS}^2.$ The constraint due to the second law
of thermodynamics is then $S_{BH} \geq S_{NS}$. For the black hole
mass larger than the solar mass  it is well satisfied as in
eq.(\ref{bhentropy}). However for the smaller  mass there can be a
cross-over $M_c$, beyond which black hole entropy becomes smaller
than the neutron star matter. We  get $M_c \sim  10^{13} g.$ It is
not clear however in what astrophysical situation this limit is
useful. For example, in neutron star matter, the degenerate
pressure is turned out to  be big enough against gravitational
collapse for $M < M_{\odot}$. One can  only speculate that it  may
give a constraint on the gravitational collapse of matter-lumps
produced in  cosmic phase transitions.
\end{thebibliography}
\end{document}